# LADIR REPORT 2010/11

# Vegetable ash as raw material in the production of glasses and enamels, for example the contemporary vegetable ashes from Burgundy, France

Ph. Colomban[a]*, A. Tournié[a], D. de Montmollin[b], L. Krainhoefner[b]

[a]Laboratoire de Dynamique, Interaction et Réactivité – UMR 7075 CNRS
Université Pierre et Marie Curie (UPMC)
2, rue Henri Dunant, 94320 Thiais, France

[b]Communaute de Taizé
71250 Taizé, France

*corresponding author
Tel : +33 1 4978 1105
Fax : +33 1 4978 1118
Philippe.colomban@glvt-cnrs.fr

**Abstract**

The powdery nature and high alkali content of vegetable ashes make them ideal raw materials to be used as modifiers of silicate compositions (glasses, enamels and ceramics). Their utilisation since ancient times is described in the literature of the history of glasses, but studies on the analyses of their composition are still limited. We discuss here the compositions of tree and shrub ashes (wattle, hawthorn, oak, green oak, olive wood, elm, poplar, apple tree, vine shoot), of plants (carex, fern, dogwood), of cereals (wheat, maize, rice), threshing waste and hay, mainly harvested in Maconnais, near Taizé (Saône-et-Loire, France), by the potter Brother D. de Montmolin. The contributions in alkali modifiers ($Na_2O$, $K_2O$), alkaline-earth (CaO, MgO) and in silica are discussed in view of the data gathered from the literature of the history of techniques used in the production of ceramics, enamels and glasses. The huge variation in composition is usually attributed to recycling and is questioned by the very broad range of compositions that we obtained in the analyses of the ashes.

# 1. INTRODUCTION

Except for fire parameters (most important the temperature and atmospheric conditions), glassmakers and potters must be able to control and enhance the melting of raw materials in order to initiate the complete melting (e.g glass, glaze, enamel) or very partial melting (e.g pottery liquid sintering) of a silicate mix. The easier solution is to play with the heterogeneity of the composition of the raw materials' and firing atmosphere to help the formation of eutectics, which has the lowest temperatures in the phase diagram. So, when the first potters fired clay in a reducing atmosphere, they obtained shards with a high densification at temperatures below 800°C, thanks to the eutectics between FeO and alkali/earth-alkali. Temperatures higher than 900-1000°C (Levin *et al.,* 1969, ibid 1975) are required in an oxidative atmosphere to obtain equivalent firing degrees. In order for glasses and enamels to homogenise, the temperature needs to be higher than the *liquidus* temperature. In any case, it is necessary to break up the raw materials to maximize the eutectic contact with the most refractory grains. The availability of raw materials consisting of fine particles is very important: hence the use of clays, limestone thermally treated (shell, chalk and shale), fine sand, flint and shocked rock crystals (heated to high temperatures and thrown into cold water to fragment or even pulvererize), also calcined shell (contribution of potash usually from vine lies), sea salt and vegetable or animal (bone) ashes. Nowadays, the ashes resulting from certain industries (smoke that condense) are used as additive in cement or in certain ceramic productions (Rawlings *et al.,* 2006; Merino *et al.,* 2005). Volcanic ashes have been used for centuries as well, (Ford & Rose, 1995).

As summarised in figure 1, several authors (Sheridan *et al.,* 2005; Artioli *et al.,* 2008) claim that the first European Bronze Age glasses were prepared from wood and plant ashes. The Mediterranean world (Brill, 1999; Tite & Shortland, 2003) would have prefered to use natron, an evaporite available in the chotts of the desert regions like the Wadi Natrum (Shortland, 2004; Shortland *et al.,* 2006), a material already used by the Egyptian dynasties for a lot of applications. The composition of natron is close to $Na_2CO_3$, $HNaCO_3$, $nH_2O$, so it is a very pure source of soda. With the end of the Roman world and the Pax Romana, the avaibility of natron became very limited and new local alkali sources searched (Cox *et al*., 1979; Wedepohl, 1997; Henderson, 2002 ; Mass, 1998), in particular the use of wood and fern ashes as claimed in the Theophilus monk's book (Cannella, 2006 ; Theophile, 1961). Whereas during the Middle Ages, halophytic plants from the chenopodiacae family (salicornia) was used (Henderson, 2002). The criteria used to attest the use of vegetable ash is the level of potassium (usually > 1.5% in soda glass); the phosphorus quantity (wood), magnesium (halophile plants) or calcium are also used clues.

Specific attention has been paid to Venetian productions at the centre of the renewal of glass item production in Europe. According to Verità and Toninato (Verità & Toninato, 1990), there are in fact two types of Venetian glasses namely "Cristallo" (high quality) and "Vitrum Blanchum" (standard quality). The cristallo is a transparent glass like rock crystal, produced from the second part of the 15[th] century. It is characterised by a high concentration of $Na_2O$ (~ 17%) and a low concentration of MgO (1.8%), $K_2O$ (2.8%) and CaO (4.9%) and is described as made from quartz pebbles and ashes imported from Syria (De Raedt *et al.,* 1999; Smit *et al*., 2005). The other group of Venetian glass, "Vitrum Blanchum", dates from the 14-16[th] century and is also a soda-rich glass, characterised by the low purity of the raw materials employed (Na$_2$O, 13.3%; MgO, 3.4%; CaO, 9.5%) (De Raedt *et al.*, 1999). During the 17$^{th}$ century, Venetian glass was used less because the famous Bohemia potassium glass and lead-rich English glass (Smit *et al.*, 2005) competed with the Venetian production.

The 19$^{th}$ century, a period of industrial revolution, is reflected in the glass industry by the importance of the technical improvement of massive productions and the control of physical properties according to glass composition.

Ternary diagrams shown in figure 2, compare the relative composition of the modifiers Na$_2$O-CaO-K$_2$O+MgO and the major oxides SiO$_2$-CaO-K$_2$O. These diagrams are useful to discriminate between the types of raw materials used for glass development (Brill, 1999; Turner, 1956; Tite & Shortland, 2003). These conclusions are broadly accepted in the literature and give the impression that the composition of the glass does not match the raw materials available at the area of their origin. This is generally explained in the literature as the re-use and mixing of glasses from diverse origins. If this proposition is well argued by the discovery of a trade of broken glass pieces coloured by rare elements (for example cobalt blue glasses) (Gratuze *et al.*, 1992; Gratuze *et al.*, 1996) from Roman period or even Phoenician (Foy, 2003; Colomban *et al.*, 2003b), the extent of this recycling is difficult to evaluate. The localisation of compositions from the literature in the main modifier diagrams (Na$_2$O-CaO-(K$_2$O+MgO), Figure 2 (Turner, 1956 ; Brill, 1999 ; Tite & Sheridan, 2003) shows that if the variety of the compositions is large, some distinct families can be distinguished from the elementary composition and also the Raman parameters characteristic of the nano-structural organisation of the silicate framework (Colomban, 2008).

The composition of the ashes vary a lot and are influenced by the nature of the selected vegetable species, the part of the plant used (foliage, branch, trunk, …), the season of harvesting and the soil where the vegetables are grown (Brill, 1999; Lambercy, 1993; de Montmollin, 1997; Oppenheim et al., 1970; Turner, 1956). This is illustrated in Figure 3, with four elementary analyses of oak samples; oak 1 and 2 analysed at few years interval, green oak and Japan oak. The difference in the used part and the washing process can explain this variability of the composition between the same types of sample. That's why the possibility to analyse an important stock of vegetable ashes made for the enamel realisation by Brother Daniel de Montmollin gives a good base of discussion to apprehend the vitreous compositions could be prepared from the vegetable ashes.

## 2. ELEMENTARY COMPOSITIONS

The vegetable ashes, sieved (washed in excess water for very alkaline or siliceous ashes, i.e. grass) then air-dried, were stocked in closed plastic bags. The wood and herbaceous vegetable ashes such as the crucifers and papilionoideae can be stocked without washing.

A lot of information on the harvest, the burning, the sieving procedures are very useful for an efficient homogenisation as described in the book "Pratique des émaux a 1300C°" (de Montmollin, 1997). Heating the samples should not be too severe in order to avoid modifier loss. The washing of straw and grass ashes needs a preliminary grinding in a jar. Several washings are generally made. This involve a loss of soluble ions and hence it one source of the composition variability of ashes of the same origin. Notice that, in a lot of cases drying leads to the formation of a hard crust at the surface due to the precipitation of soluble salts (efflorescence).

The quantity of ash obtained relative to the harvest product (dry) used, vary between around 0.5% (oak, beech timber), 1% (branch, shrub, fruit tree, conifer), 3-5% (fern, straw), the maximum (18%) is obtained for horsetail. We can deduce that the percentage of ash increases with the silica content.

The colours of the different samples vary from light grey to black according to the quantity of carbon and coloured oxides (Table 1). Thermogravitic analyses and the fire loss at 150°C have been done, because these powders are very hygroscopic, in order to be able to calculate the elementary compositions on the "dry" product. Usually the weight lost on ignition varies from 10% to 30% according to the type of ashes and their preparation method.

The elementary content of S, Si, Al, Fe, Ti, Ca, Mg, K, Na, P, Mn and C were determined by the Analysis Central service of the CNRS (Vernaison, France) using ICP-AES (Thermo-Fisher instrument) on products dried at 150°C, which certify a water residual content less than a few percentages. The results have been converted to oxide mass % equivalent (Table 1). The carbon contents have been measured on ash lists, which have gradual coloration levels: the range of carbon content is a few percentages for the grey samples until close to 10% for the "black" ashes.

We can classify from the manufacturers point of view, the ashes into two groups namely acidic ashes, which are very rich in silica (grass ashes, horsetails, carex and ferns), and those rich in basic oxides which are modifiers (mainly wood). Several analyses show that the silica content increases from the trunk to the leaves, for example the percentage is multiplied by 10 for the beech. The analyses of silica content confirm that rice is a very pure source of silica (close to 94%), moreover it is the only vegetable ash, to our knowledge, industrially used (Bondioli *et al.*, 2007). A very high content in silica as well for maize 73, dogwood and carex (~ 70-80%), for hay and wheat (50-65%, the grain is very low in silica!) and between 45 to 55% for the poplar, fern and maize. Wattle wood as well as some hay, oak and wood like the poplar, yield significant quantities aluminium oxide (5-7%). Some vines can reach close to 30% of aluminium oxide (Lambercy, 1993). The vine shoot, carex, hay, wheat and wattle supply iron oxide (2.5%-3%). The contribution in calcium oxide is prevailing in the wood: 85% for the green oak, 70% for olive wood, 55-63% for the others trees, the low contents is observed for the vine shoot (45%), hawthorn and the poplar (~30%), the hay and fern (~20%), the wattle, some plants (carex, dogwood) and cereals (≤15%). So, wood ash yields a lot more calcium than bone ash (~50-55%). The variability of potassium contents are less than 15% for most of ashes except some wheat, oak and fern (15-20%), wattle and hawthorn (25 to 30%) and red clover (35%, (Lambercy, 1993)). The record of potassium content seem to be obtained by the potato tuber ashes with close to 60%, the balance is made up mainly by $P_2O_5$ (16%) and MgO (5%) (Lambercy, 1993). Otherwise, the $P_2O_5$ oxide content is a maximum for the threshing waste around 20%, for some wheat around 14% and reach 5-10% for most shrubs and tree ashes (wattle, apple tree, elm, hawthorn, vine shoot), few plants (carex, dogwood), hay and maize. The range of the MgO content is quite small (2-5%) except for wattle (9%), fern (8%) and hawthorn, some wheat and threshing waste (5-6%).

The variability of phosphorus content shows that a small percentage of this element is not a proof that vegetable ashes were not used, contrarily to a common assertion. The phosphorus content of hay and cedar depends on the ripeness of the grains and is at a maximum if they are ripe. Vegetable ashes seem to be an important calcium source. Obviously, the contribution of sodium is negligible with continental vegetation.

The ash colours inform on the chromophore oxide present: purple if manganese, grey if carbon, brown to black if presence of manganese and iron and/or carbon mix.

## 3. RAW MATERIALS IDENTIFICATION

Figure 4 summarizes the composition area relative to modifier and modifier/silica according to the vegetable ashes of Table 1. In an easy way, it can be seen that if wood is used a composition rich in calcium is obtained, while grass or cereals leads to compositions rich in potassium. However the effect of the potassium modifier is limited when the content in silica is important.

Because of the very low level of sodium in most of the vegetable ashes, it is impossible to produce, without the supplementary contribution of sodium, glasses with the compositions in figure 2. Thus, glasses made from "continental" ashes also needed the addition of other raw materials such as river sand containing sodium feldspars (for examples albite, sanidine) and/or sea salt NaCl or sodium sulphate. The analysis of ancient texts brings some partial answers. The Cannella's report (Cannella, 2006) which analyse beyond Jean d'Outremeuse manuscript, shows in the entirety of glass recipes texts the regularly use of salt or alkali powder (p. 126-129). The use of sulphate seems to be rejected because to incorporate sulphate into a melted glass is difficult. Moreover the real nature of reagents seems arguable; particular those called "calcine scale or soda" (sic.). Jean d'Outremeuse indicates the use of "very white alkali salt" (Cannella, 2006, p.132) obtained from plants. Then it seems established that the coastal plant ashes, rich in soda were available and so in complement of "continental" ashes let to reach the whole of compositions named "continental plants" (figure 2a), area manifestly more broad that shows in figure 3a obtained with the ashes from table 1. Cannella report that according to Verità (Verità, 1991c, p133) the soda is imported from the Middle East (Syria-Egypt) at Venice (*alume catino, cenere de Levante, cenere di Soria*) and consist of *salsola soda* or kali ashes.

Figure 5 indicates the locations of German and French cathedral stained glass window compositions (Sterpenich, 1998) as well as the liquidus and temperatures from the phase diagram (Morey *et al.*, 1930). These composition areas are coherent with other work (Schalm *et al.,* 2004). The compositions of German stained glass windows dated from $14^{th}$ century rich in calcium and that French/German dated from 7-$14^{th}$ century rich in potassium, match very well with a preparation from "continental" ashes while it is obvious that the stained glass windows of Rouen dated from 7-$9^{th}$ century rich in sodium were made with soda (salt or coastal plant ashes or imported salt lake).

## 4. NON-DESTRUCTIVE ANALYSE OF GLASSES

The identification of composition types according to sampling is not possible for every sample. An efficient alternative is now possible thanks to Raman spectroscopy (Colomban, 2008), a technique that can be used on site, for example on stained glass windows in the Sainte-Chapelle (Colomban & Tournié, 2007), on items which can't leave their conservation place (Ricciardi *et al*., 2009a; Kirmizi *et al.*, 2010a; ibid,

2010b). This identification is obtained from the analyses of the silicate network nanostructure $[Si-O-Si-O-Si]_n$ polymer which depends mainly on the composition, in particular the modifier content (Na, K, Ca) which determines the polymerisation degree of the silicate network (Colomban, 2008; Colomban *et al.*, 2006a), and incidentally the thermal history. The Raman spectrum is a function of the degree of covalency of the tetrahedral network and for example the silica/alumina ratio. Then if the glass or enamel compositions can be discriminated by their modifier content (and their nature) or polymerisation index (Colomban, 2003a; ibid, 2003b) and by their ratio of silica/alumina, Raman spectroscopy will be efficient (Colomban *et al.*, 2006a). Figure 6 presents the classifications made from the literature data (Colomban *et al.*, 2006b; Lagabrielle & Velde, 2003; Schalm, *et al.*, 2004; Sterpenich, 1998; Tournié *et al.*, 2006) for the stained glass windows of religious buildings from France, Belgium, Netherlands and Germany. These diagrams prove that during the technology renewal period (14$^{th}$ century in Germany, 19$^{th}$ century in Europe, America and Japan) the variability of the compositions is very large and so Raman spectroscopy is able to discriminate very well between the different kinds of stained glass windows. It is also possible for other productions because their modifier content and/or the covalence degree change (Colomban *et al.,* 2006a; ibid, 2006b).

From the Raman spectrum, the maximum of the position of the bending and stretching massifs (respectively named $\delta_{max}$ SiO$_4$ and $\nu_{max}$ SiO$_4$) are well established tools for the classification of amorphous silicates (Colomban & Paulsen, 2005; Colomban, 2006b; Colomban *et al.*, 2004; Colomban & Tournié, 2007). Figure 7 classify in function $\delta_{max}$ SiO$_4$ and $\nu_{max}$ SiO$_4$ a very large corpus of glass constituted of stained glass windows (13-19$^{th}$ centuries, previously used in figure 6, Colomban *et al.*, 2006b), glass objects mostly enamelled of various origins (France, Italy, Netherland, Spain) from 16-19$^{th}$ centuries studied on site at the Ceramics National Museum at Sèvres (France). Finally, painted enamel items, which belong to Limoges production, from 16-19$^{th}$ centuries, studied as well on site in the storage room of the Museum of Arts Deco in Paris (Kirmizi *et al.,* 2010b). The figure 7 shows a dispersion of the composition in groups which are relatively well defined. For each period, 16-17$^{th}$ c. or 18-19$^{th}$ c. in the same group (Sèvres, glass windows or Limoges) we can find different composition, which seems due to the very wide diversity of vegetable ash compositions.

## 5. CONCLUSION

The experimental determination of the composition of different kinds of "continental" plant ashes shows that the variability of the compositions is very large. The very low sodium content confirms that for the preparation of sodium glass and even potassium-sodium mixed glasses raw materials rich in sodium (imported maritime plant ashes, natron or sea salt). The differences of content in alkali/earth-alkali from the use of different kinds of raw materials are sources from the effectiveness of Raman analyse, non-destructive technique could be used on site with mobile instrument.

# Références

| | Sample | Reference | Dry residue / coulour | C | SO$_4$ | SiO$_2$ | Al$_2$O$_3$ | Fe$_2$O$_3$ | TiO$_2$ | CaO | MgO | K$_2$O | Na$_2$O | P$_2$O$_5$ | Total with C | total without C |
|---|---|---|---|---|---|---|---|---|---|---|---|---|---|---|---|---|
| Tree shrub | Wattle | ACAC | Beige | 8,58 | 4,46 | 22,31 | 7,25 | 2,45 | 0,47 | 16,53 | 9,25 | 25,34 | 1,02 | 10,92 | 108,58 | 100 |
| | | SCA 05007883 | | 8,54 | 5,29 | 23,85 | 7,39 | 2,49 | 0,47 | 15,63 | 9,13 | 24,94 | 0,6 | 10,22 | 108,55 | 100,01 |
| | Hawthorn | AUBEP | Gris clair | | 3,55 | 22,34 | 1,71 | 0,76 | 0,11 | 31,39 | 5,33 | 27,45 | 0,89 | 6,47 | 100 | 100 |
| | | SCA 05007883 | | | 0 | 22,12 | 1,81 | 0,84 | 0,12 | 32,71 | 5,69 | 29 | 0,94 | 6,76 | 99,99 | 99,99 |
| | Olive wood | OLIV | Beige | | 1,63 | 6,21 | 2,95 | 0,7 | 0,14 | 70,62 | 4,7 | 9,32 | 0,58 | 3,16 | 100,01 | 100,01 |
| | Elm | ORME | Gris clair | | 2,91 | 16,25 | 1,21 | 0,71 | 0,05 | 58,89 | 4,89 | 8,89 | 0,4 | 5,81 | 100,01 | 100,01 |
| | Poplar | PEUPL | Gris clair | | 5,85 | 46,01 | 4,76 | 1,32 | 0,18 | 27,3 | 1,9 | 10,15 | 0,58 | 1,95 | 100 | 100 |
| | Apple tree | POM | Beige | | 0,57 | 12,2 | 1,81 | 1,13 | 0,08 | 62,57 | 4,6 | 11,92 | 0,11 | 5 | 99,99 | 99,99 |
| | | SCA 05007883 | | | 0 | 11,21 | 2,17 | 1,27 | 0,08 | 63,54 | 4,6 | 11,85 | 0,11 | 5,17 | 100 | 100 |
| | Oak | CHENE | Beige marron | | 5,07 | 13,42 | 1,52 | 1,43 | 0,11 | 54,69 | 2,07 | 18,47 | 0,35 | 2,86 | 99,99 | 99,99 |
| | Green oak | CHEVT | 61.50 Beige | | 0 | 2,44 | 1,32 | 0,34 | 0,05 | 84,5 | 1,83 | 7,52 | 0,09 | 1,92 | 100,01 | 100,01 |
| | | SCA 05007883 | | | 0 | 2,87 | 1,27 | 0,3 | 0,05 | 84,33 | 1,81 | 7,47 | 0,12 | 1,78 | 100 | 100 |
| | Vine shoot | SARMT | 72.00 Beige marron | | 1,87 | 20 | 4,2 | 3,01 | 0,25 | 46,39 | 4,82 | 14,43 | 0,23 | 4,8 | 100 | 100 |
| plant | Carex | CAREX | Gris foncé | 3,89 | 0 | 67,42 | 2,62 | 2,31 | 0,08 | 9,48 | 3,67 | 6,77 | 0 | 7,64 | 103,88 | 99,99 |
| | Fern | FOUGR | Gris | | 0 | 48,87 | 1,29 | 1,63 | 0,1 | 18,71 | 8,21 | 16,68 | 0,92 | 3,6 | 100,01 | 100,01 |
| | Dogwood 71 | QUE71 | Gris foncé | 9,46 | 0 | 68,76 | 3,23 | 1,04 | 0,14 | 3,46 | 4,96 | 11,34 | 0,2 | 6,86 | 109,45 | 99,99 |
| Cereal | Wheat 04 | BLE04 | Noir | | 0,46 | 65,15 | 1,01 | 0,91 | 0,06 | 8,69 | 3,34 | 16,67 | 0,21 | 3,5 | 100 | 100 |
| | Wheat 76 | BLE76 | Gris clair | | 0 | 51,41 | 2,89 | 2,28 | 1,29 | 11,19 | 5,8 | 9,93 | 1,32 | 13,89 | 100 | 100 |
| | Maize 03-CH | MA3CH | Gris foncé | 7,86 | 0 | 47,72 | 3,01 | 0,52 | 0,34 | 40,43 | 1,91 | 4,42 | 0,31 | 1,33 | 107,85 | 99,99 |
| | Maize 03-StR | MA3SR | Gris foncé | | 0 | 53,18 | 4 | 1,49 | 0,23 | 15,07 | 4,11 | 14,14 | 0,31 | 7,48 | 100,01 | 100,01 |
| | | SCA 05007883 | | | 0 | 52,03 | 4,17 | 1,56 | 0,23 | 15,14 | 4,28 | 14,5 | 0,4 | 7,69 | 100 | 100 |
| | Maize 73 | MA73 | 91.00 Gris | | 0 | 76,98 | 3,33 | 0,77 | 0,14 | 6,44 | 2,35 | 6,83 | 0,3 | 2,87 | 100,01 | 100,01 |
| | Rice | RIZ | Gris foncé | 0,37 | 0 | 93,81 | 0,22 | 0,2 | 0 | 1,09 | 0,88 | 2,74 | 0,11 | 0,95 | 100,37 | 100 |
| Hay | Hay 70 | FOIN70 | Gris | | 0 | 48,61 | 4,91 | 2 | 0,24 | 20,68 | 4,24 | 12,6 | 0,85 | 5,87 | 100 | 100 |
| | Hay 78 | FOIN78 | Gris | | 0 | 52,06 | 6,3 | 2,98 | 0,31 | 17,12 | 3,52 | 10,78 | 1,05 | 5,89 | 100,01 | 100,01 |
| Waste | Threshing waste | DECBA | Gris | 3,7 | | 37,56 | 2,59 | 1,63 | 0,1 | 16,41 | 5,54 | 10,43 | 0,34 | 21,71 | 100,01 | 100,01 |
| | | SCA 05007883 | | | 0 | 39,49 | 2,74 | 1,63 | 0,1 | 17,23 | 5,69 | 10,82 | 0,35 | 21,95 | 100 | 100 |

Table 1: Elementary compositions of vegetable ashes from Burgundy (France), analysed by ICP-AES

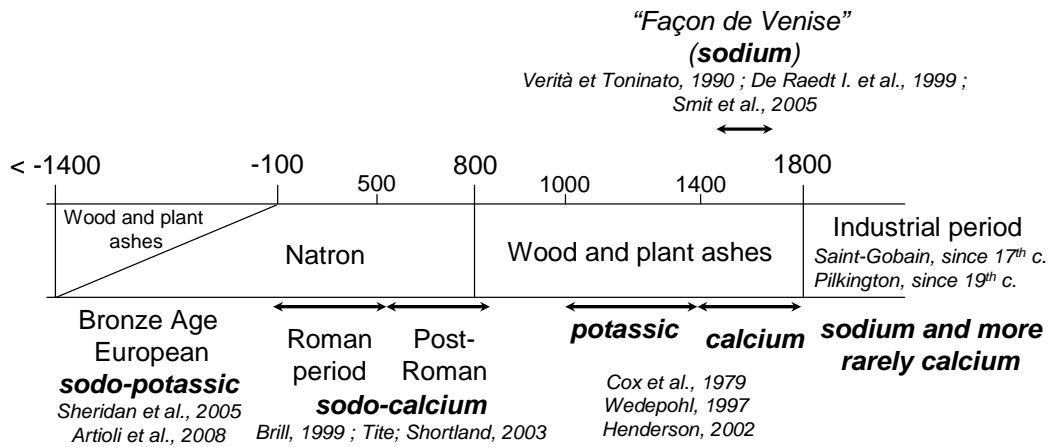

*Figure 1*: Main sources of modifiers for the glasses.

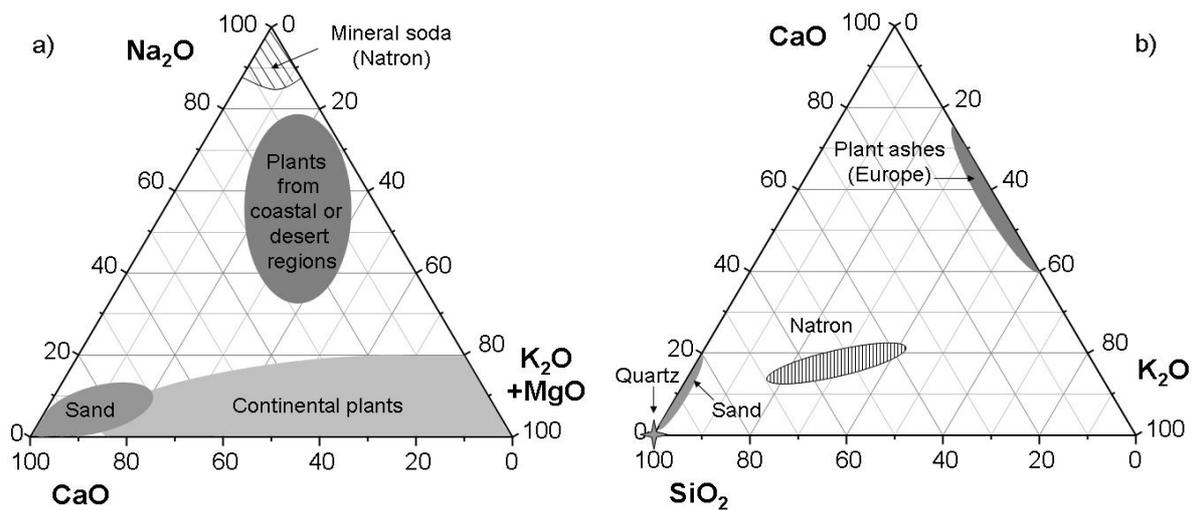

*Figure 2*: Ternary diagrams of different types of modifiers and formers a) $Na_2O$-$CaO$-($K_2O$+$MgO$) and b) $CaO$-$SiO_2$-$K_2O$ create from the literature data (Turner, 1956 ; Brill, 1999 ; Tite & Shortland, 2003).

**Figure 3**

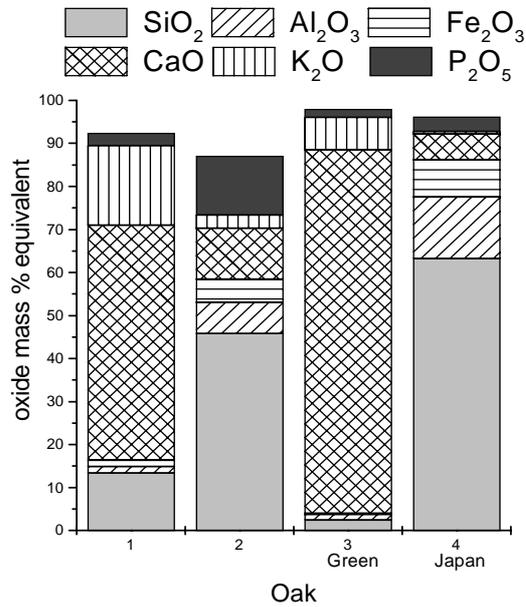

*Figure 3:* Column graphic compares different oak elementary compositions.

**Figure 4**

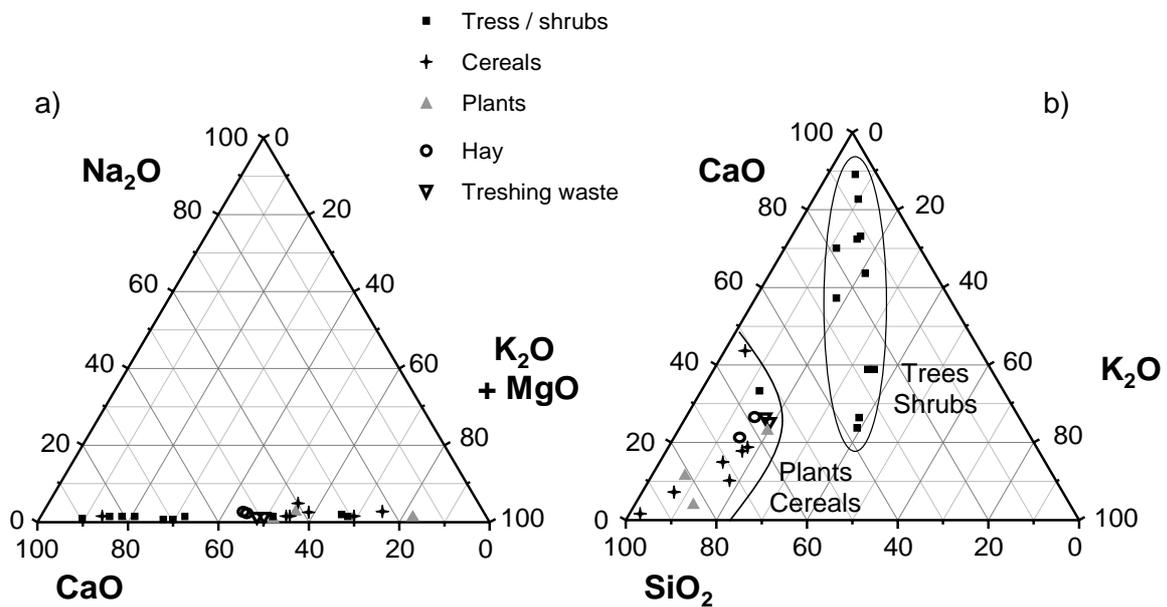

*Figure 4*: Ternary diagrams a) Na$_2$O-CaO-(K$_2$O+MgO) and b) CaO-SiO$_2$-K$_2$O of different types of vegetable ashes collect in the Taizé region (Saône-et-Loire, France) make up with trees/shrubs (wattle, hawthorn, oak, olive wood, elm, …), cereals (wheat, maize, rice) and plants (carex, fern, dogwood).

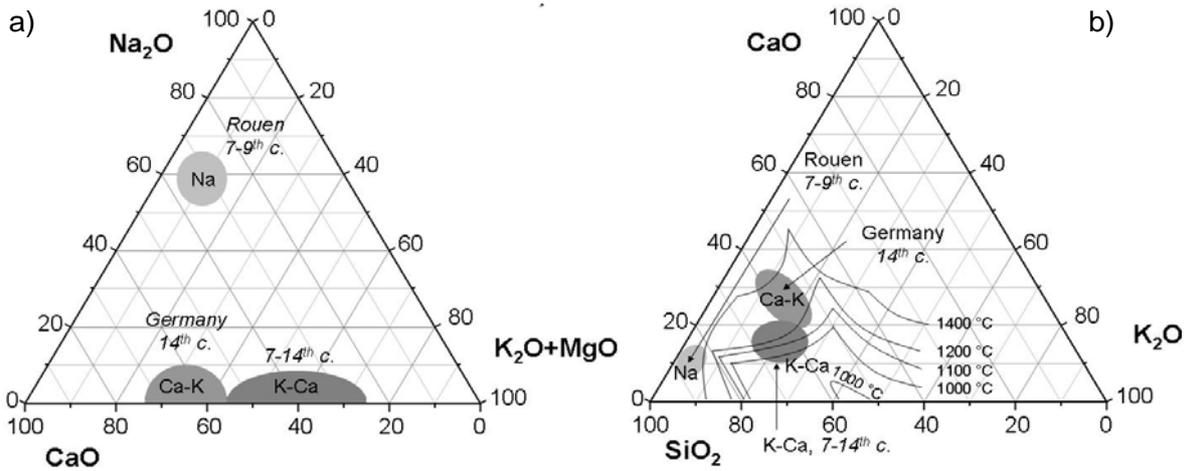

*Figure 5:* Comparison in the ternary diagrams a) $Na_2O$-$CaO$-($K_2O$+$MgO$) and b) $CaO$-$SiO_2$-$K_2O$ of compositions of German and French cathedral stained glass windows ($7^{th}$-$14^{th}$ centuries, Sterpenich, 1998). The liquidus are transferred from the ternary phase diagram of Morey (Morey *et al.*, 1930).

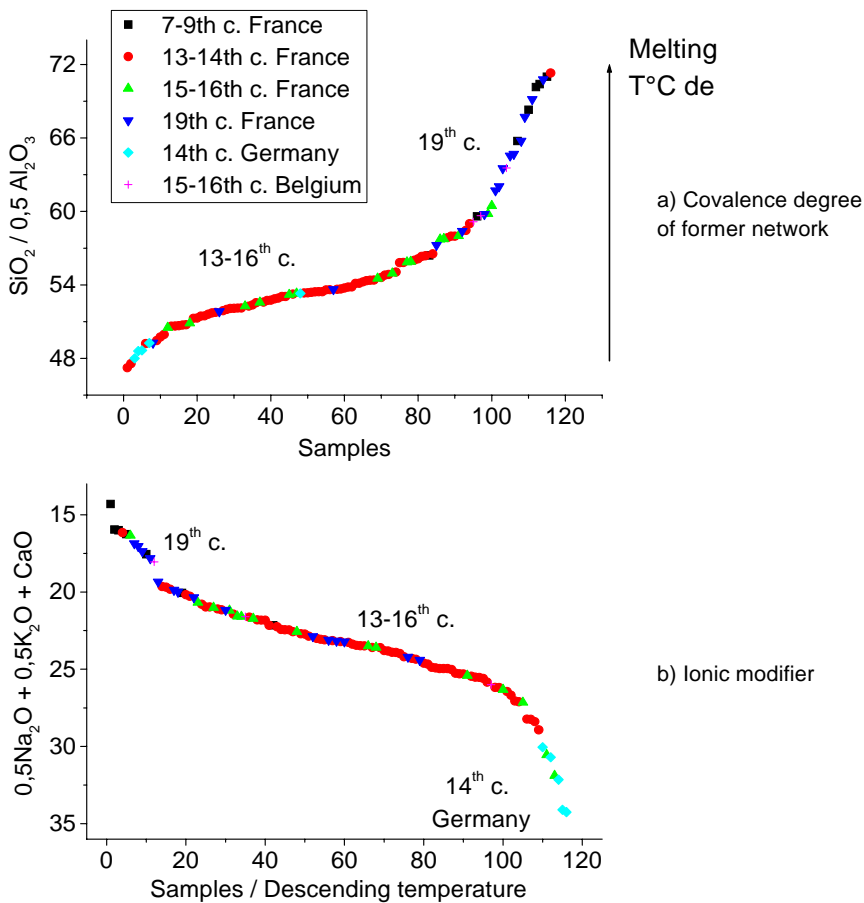

*Figure 6*: Classification of stained glass windows from a) covalence index ($SiO_2/0.5Al_2O_3$) and b) ionic modifier content ($0.5Na_2O+0.5K_2O+CaO$).

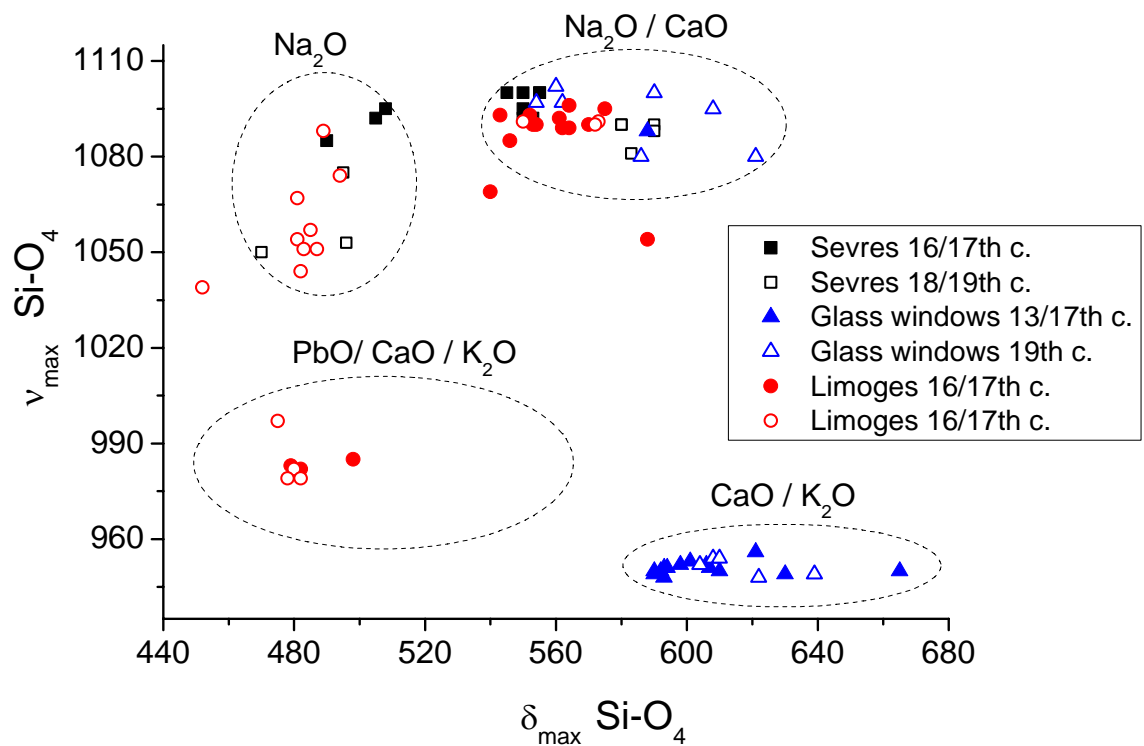

*Figure 7:* Plot of the Si-O stretching maximum values versus bending maximum values for a very large corpus of glass; stained glass windows, glass objects mostly belong to the Ceramic National Museum at Sèvres and the Museum of Arts Deco at Paris (Limoges production).